\documentclass[letterpaper, 10 pt, conference]{ieeeconf}  

\IEEEoverridecommandlockouts                              

\usepackage{graphics} 
\usepackage{epsfig} 
\usepackage{mathptmx} 
\usepackage{times} 
\usepackage{amsmath} 
\usepackage{amssymb}  
\usepackage{bm}
\usepackage{algorithm}
\usepackage{multirow}
\usepackage{verbatim}
\usepackage{graphicx}
\usepackage{hyperref}
\usepackage{graphicx,fancyhdr}
\usepackage{listings}

\usepackage{booktabs} 

\usepackage{subfigure}
\usepackage{color}

\title{Filtered 2D Contour-Based Reconstruction of 3D STL Models from CT-DICOM Images}

\author{K.Punnam Chandar$^{1}$, Y.Ravi Kumar$^{2}$
\thanks{$^{1}$Kakatiya University, $^{2}$National Institute of Technology, Warangal}
} 
\begin{document}

\maketitle
\thispagestyle{empty}
\pagestyle{empty}

\begin{abstract}
Reconstructing a 3D Stereo-lithography (STL) Model from 2D Contours of scanned structure in Digital Imaging and Communication in Medicine (DICOM) images is crucial to understand the geometry and deformity. Computed Tomography (CT) images are processed to enhance the contrast, reduce the noise followed by smoothing. The processed CT images are segmented using thresholding technique.  2D contour data points are extracted from segmented CT images and are used to construct 3D STL Models. The 2D contour data points may contain outliers as a result of segmentation of low resolution images and the geometry of the constructed 3D structure deviate from the actual. To cope with the imperfections in segmentation process, in this work we propose to use filtered 2D contour data points to reconstruct 3D STL Model. The filtered 2D contour points of each image are delaunay triangulated and joined layer-by-layer to reconstruct the 3D STL model. The 3D STL Model reconstruction is verified on i) 2D Data points of basic shapes and ii) Region of Interest (ROI) of human pelvic bone and are presented as case studies. The 3D STL model constructed from 2D contour data points of ROI of segmented pelvic bone with and without filtering are presented. The 3D STL model reconstructed from filtered 2D data points improved the geometry of model compared to the model reconstructed without filtering 2D data points. 

\end{abstract}

\section{INTRODUCTION}
        \label{sec:introduction}
        Computed Tomography (CT) images provide detailed information to clinicians than conventional X-rays. A fixed number of 2D slices are collected and are digitally stacked to virtually view region of interest of internal structure of human body in voxel model. The voxel models are not suitable for bio-medical applications owing to their low resolution and inconvenient modification \cite{young2008efficient}. 3D medical modeling procedure usually aims to convert the original voxel model into another kind of 3D solid model suitable for bio-medical applications. Medical images in 3D solid model allows doctors for easier identification of abnormality/deformity in internal structure 
\cite{preim2007visualization,karatas2014three, kanumilli2024advancements}. All reconstructed 3D solid models can be converted to rapid prototyping (RP) models to process them in two phases: virtual(modeling and simulating) \cite{ahn2006rapid,willis2007rapid, marovic1998visualization, pflesser1995towards} and physical (fabrication) \cite{rosa2004rapid,wang2010stl, manmadhachary2016improve, negi2014basics}. 

To aid in virtual phase, there are three methods  that can be applied to reconstruct a 3D solid model for bio-medical applications: polygonal model, parametric surface model and third 2D contour detection method. Polygonal Model is via voxel to stack and construct the model by using the marching-cube algorithm \cite{ma2001rapid,rajon2003marching}. Parametric surface method involves a swept from the contours of each layer. The reconstructed surface can be then lofted from these contours 
\cite{choi1991surface,lee2002feature, meyers1992surfaces, shang2018closed}. 2D contour detection method, each slice of CT image is separately processed to extract the pile of contours and the 2D contour points are used to construct the delaunay triangulation and added with successive layers mesh by parallel polygons to form 3D model
\cite{wang2010stl, maksimovic2000computed, choi1988triangulation, schumaker1993triangulations}.

 Each of these methods has certain limitations/disadvantages. With the first method, the shape modification is usually tricky and time consuming, which makes it unsuitable for RP applications 
\cite{sisias2002algorithms, lin2007voxelization}. The parametric surface method requires drawing of curve model in CAD and is extremely complicated. Since the spline must first be constructed before modeling can take place
\cite{leong1996study}. The 2D contour detection method suffers from drawbacks in contour detection for each CT slice and file errors in the construction and use of delaunay triangulation
\cite{stroud2000stl, wu2006enhanced, melchels2010review}.

The present work, we propose an enhanced process for converting a CT medical images to a 3D delanunay triangulation mesh model for easing the modeling and simulation. The process of reconstructing the medical 3D mesh model can be achieved in two phases, viz., i) medical image processing and segmentation and ii) 3D reconstruction of STL mesh. Medical image processing is achieved through traditional work flow which serves to provide image enhancement followed by image segmentation. The 2D Contour data points are extracted from segmented images of each slice and are subjected to filtering operation to reduce outliers. The filtered 2D Contour data points are sorted out and delaunay triangulation is performed and adjacent layers are connected  layer-by layer using extra polygon layer to reconstruct 3D STL model.
        
\section{CT Image Processing}
        \label{sec:ct image processing}
\begin{figure}[t]
\centering
\includegraphics[width=.40\textwidth]{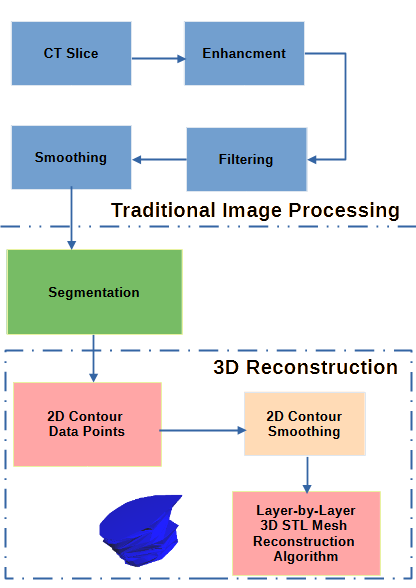}
\caption{Flow Chart of CT Images to 3D STL Model}
\label{fig:pipeline}
\end{figure}

        The CT scan images are to be processed before extracting the contour of the internal structure in each slice. First, enhancement is performed followed by filtering and then smoothing before proceeding for segmentation. Flow chart of the CT images to 3D STL Model reconstruction is shown in Figure.\ref{fig:pipeline}

\subsection{Traditional Image Processing}
\begin{figure}[t]
\centering
\includegraphics[width=.40\textwidth]{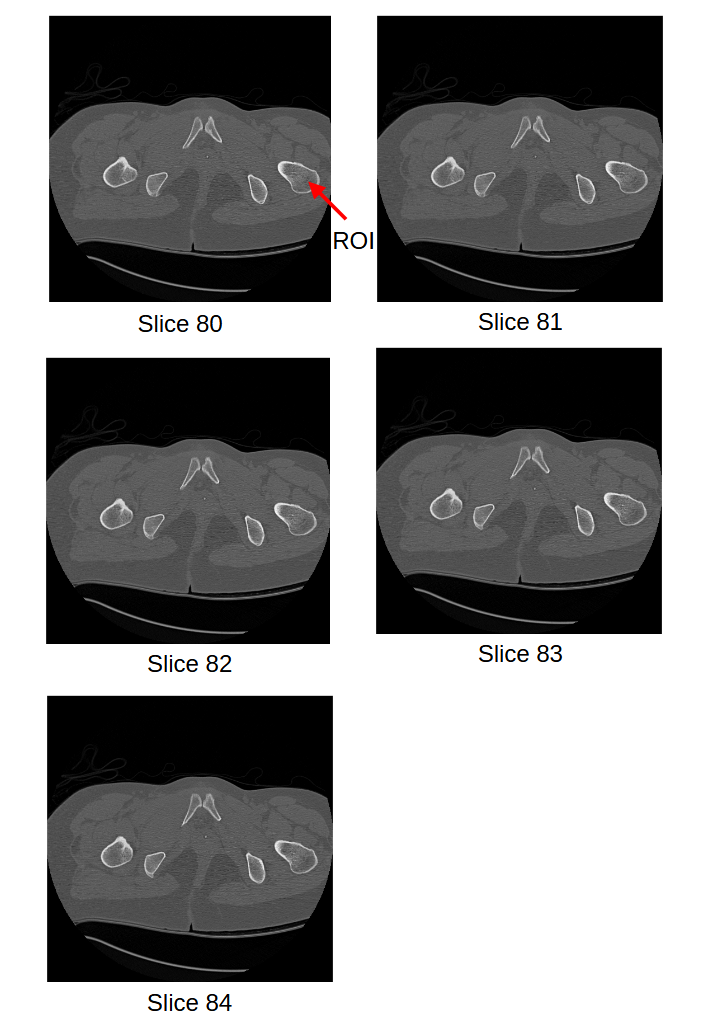}
\caption{CT Images Slices 80-84 with Region of Interest}
\label{fig:sample CT}
\end{figure}

CT images assist the clinical practitioners to assesses the internal structure for deformity when met with serious injuries. CT images captures the intensity variations of different internal body parts like bones, muscles, fat, organs and blood vessels relative to the water in Hounsfield Unit \cite{exampleWebsite}. The HU are mapped linearly to the gray scale range [0, 255].  CT Scanning captures human body in incremental scanning and provides up to 350 slices depending on the thickness of the slice or resolution of the CT scanning machine. The practicing clinicians needs to analyze each slice manually to precisely locate the deformity or injury in the internal structure. Manually analyzing slice-by-slice is a tedious task and the 3D structure need to be constructed in his imagination based on his learning and experience. Most important internal organ that will be assessed in all these CT slices is bone. Segmenting the bone in all these slices and
reconstructing the 3D Structure will assist the clinicians in saving time and enough room for other medical procedures. Sample CT scan images 
\cite{databaseKey} of pelvic bone from slice 80 to 84 with region of interest Pelvic bone femoral body is shown in Figure~\ref{fig:sample CT}.

To process the CT scan images to reconstruct 3D Model of internal bone structure, the images need to be processed in different phases viz, Traditional Image Processing (TIP) phase, Segmentation phase, Contour extraction and filtering phase and finally layer-by- layer reconstruction of 3D STL Medical Model of the region of interest of the internal structure.

Complex internal bone structure, speckle noise captured during scanning and intensity inhomogeneities need to be processed before performing the task of segmentation.  The TIP Phase shown in Figure~\ref{fig:pipeline},aims to handle these challenges before proceeding for segmentation phase 
\cite{wang2013feature,chandar2016segmentation}. In TIP phase, first the CT image slice is enhanced using power-law transformation given in equation~\ref{eq:scaling}, the variable \(s\) depends on \(c\), \(r\), and \(\gamma\). 

\begin{equation}
s = c \cdot r^{\gamma}
\label{eq:scaling}
\end{equation}

In equation~\ref{eq:scaling}, the input \(r\) is CT image intensity value in the range [0, 255], where intensity 0 refers to dark and 255 white. To improve the details in the dark regions the value of \(\gamma\) is to be specified. Lower values of \(\gamma\) results in enhancement of CT images with dominant presence of speckle noise. Median filtering will follow the power-law transformation, as the median filter performs well in the presence of speckle noise. A Median filter Mask size of 9x9 reduces speckle noise considerably. To process, any residual speckle noise present in the form of discontinuities smoothing filter of mask size 9x9 is used. The parameters considered in MATLAB simulation for pre-processing are c=1 and \(\gamma\)=0.3.

\subsection{Segmentation}
\begin{figure*}[h]
\centering
\includegraphics[width=\textwidth]{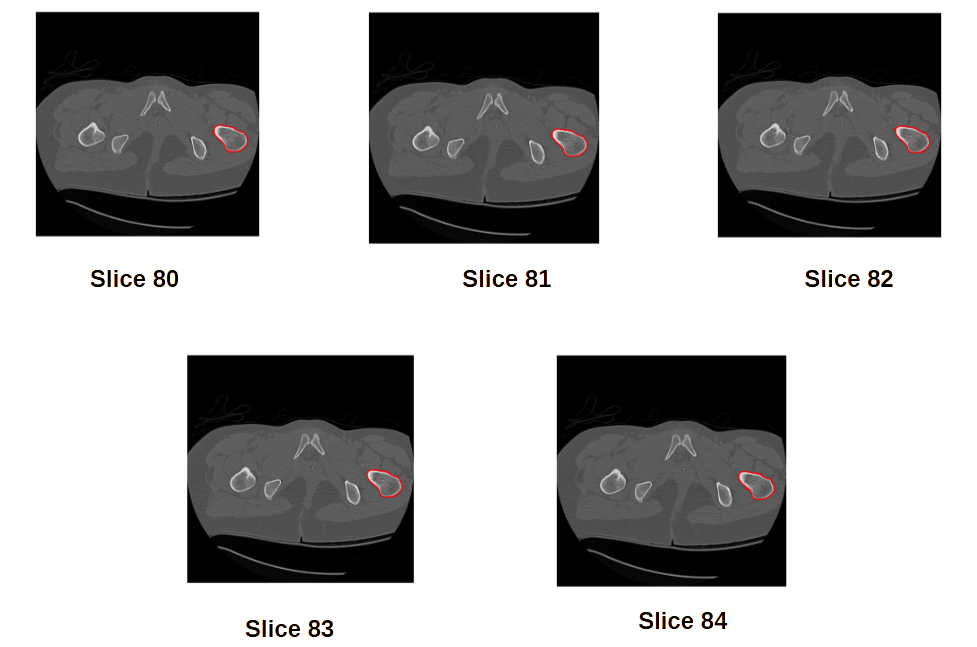}
\caption{Contours of ROI}
\label{fig:contours}
\end{figure*}

After the traditional work flow, segmentation of the internal structure is performed on each slice of region of interest. To perform the task of image segmentation, each slice of CT Image is subjected to thresholding and followed by morphological operations, the boundaries of ROI are extracted. The results of segmentation on slice 80 to slice 84 are shown in Figure~\ref{fig:contours} for a threshold value above 400.

\subsection{Filtering 2D Contours}
\begin{figure}[t]
\centering
\includegraphics[width=.40\textwidth]{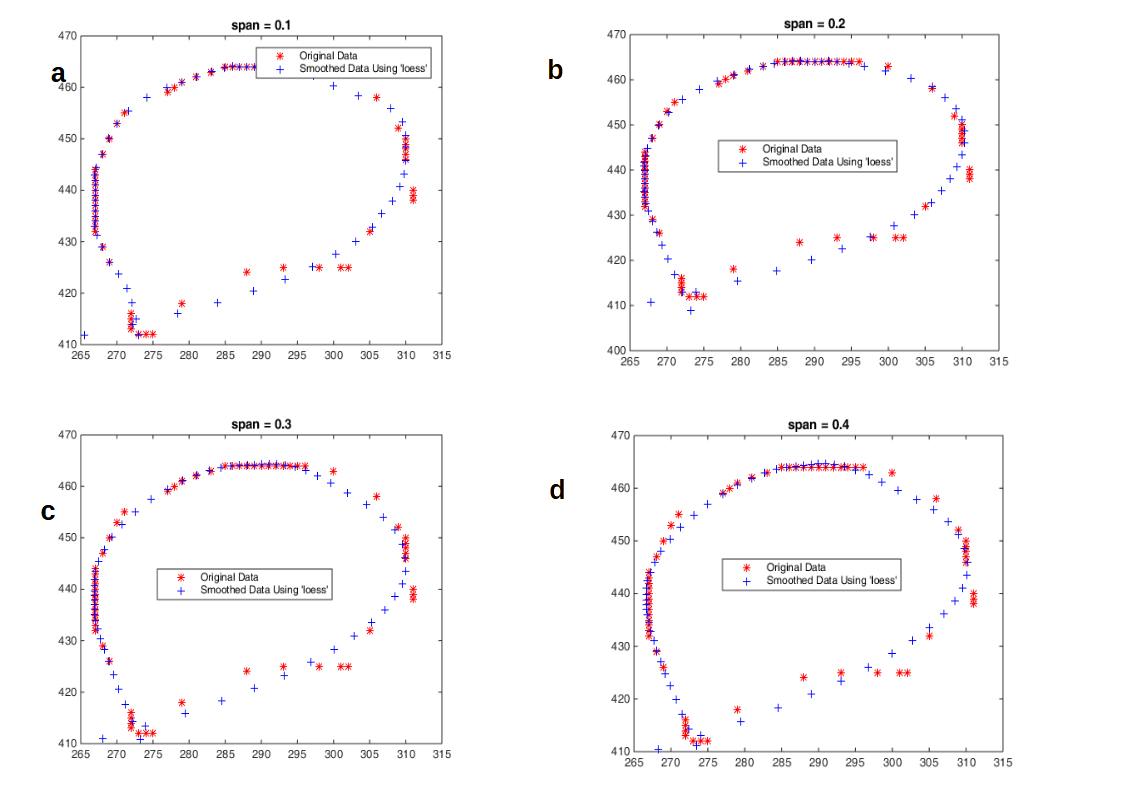}
\caption{Smoothing of contours with varying span values.}
\label{fig:smoothing}
\end{figure}

The extracted contours from segmented low resolution CT images are corrupted with the presence of outliers as shown in Figure~\ref{fig:contours}. Before proceeding for 3D STL Model reconstruction, the contours need to be filtered. Filtering is performed on \(x\) and \(y\) coordinates separately using Moving Average Filter in MATLAB. The number of data points considered for smoothing is specified using parameter ‘span’ and is a scalar value in the range (0, 1), with smoothing method Local regression using weighted linear least squares and a 2nd degree polynomial model. The snippet of the MATLAB code is given below:

\begin{lstlisting}[language=Matlab]
x = smooth(x,span,method)
y = smooth(y, span, method)
\end{lstlisting}

The 2D Contour Data points of slice 80 are smoothed with varying scalar values of ‘span’ and the results are presented in Fig.4. Contour smoothing with ‘0.1 span 0.4 in steps of 0.1, is shown in Figure~\ref{fig:smoothing}, (a), (b), (c), (d), where smoothing results are more sensitive to outliers and with ‘span = 0.1’, there is a smooth contour with good outlier rejection.
 
\begin{figure}[t]
\centering
\includegraphics[width=.40\textwidth]{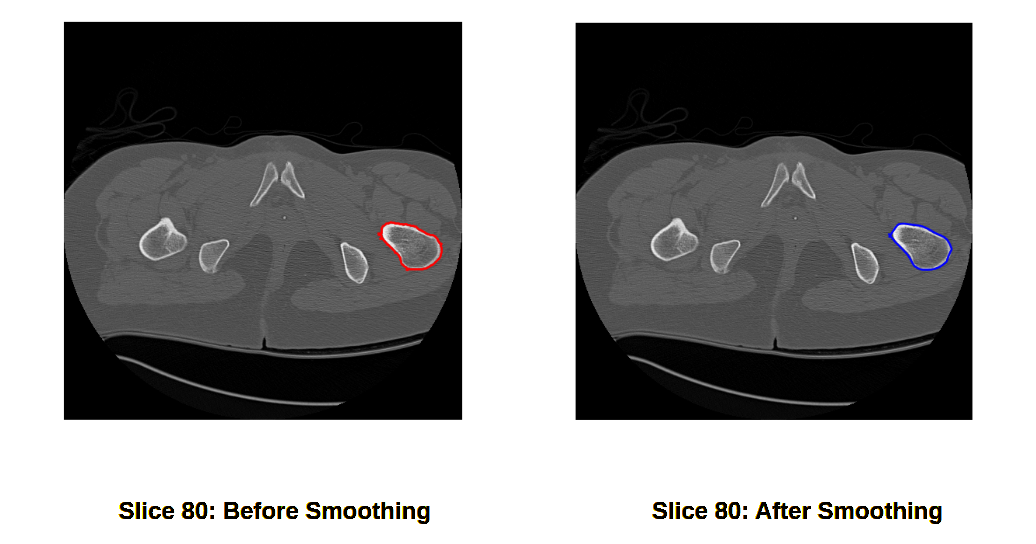}
\caption{2D contour data plot of slice 80 before filtering and after filtering.}
\label{fig:filter80}
\end{figure}

\begin{figure}[t]
\centering
\includegraphics[width=.40\textwidth]{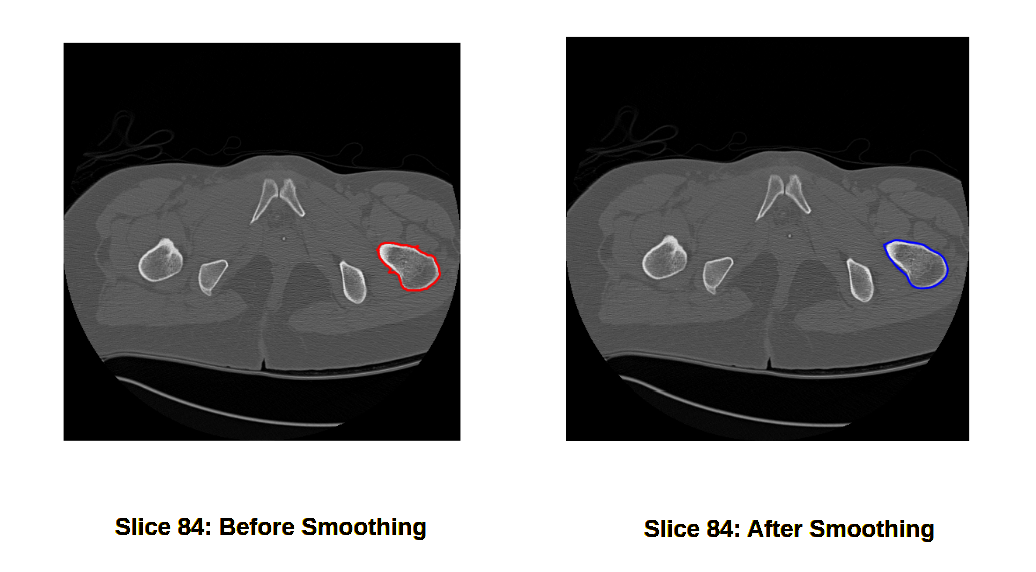}
\caption{2D contour data plot of slice 84 before filtering and after filtering.}
\label{fig:filter84}
\end{figure}

The 2D Contour Data points of CT Slices 80-84 are filtered with moving average filter with span=0.1 and the results are shown for slice 80 in Figure~\ref{fig:filter80} and for slice 84 in Figure~\ref{fig:filter84}. 

\section{3D STL Mesh Reconstruction}
        \label{sec:method}
        \begin{figure*}[h]
\centering
\includegraphics[width=.50\textwidth]{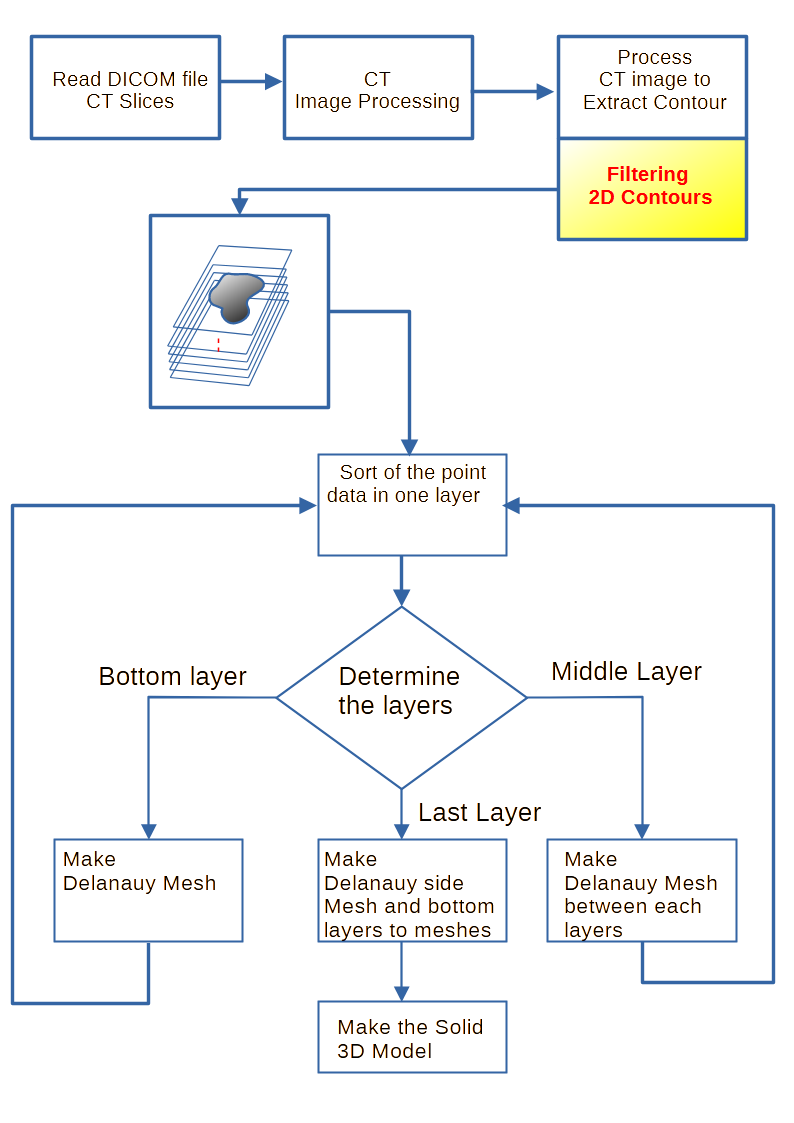}
\caption{Flow chart for CT Images to 3D STL Mesh generation.}
\label{fig:flowchart}
\end{figure*}

Flow chart of Layer-by-Layer 2D contour data points to 3D STL Mesh generation algorithm is shown in Figure~\ref{fig:flowchart}. The algorithm presented in this paper is enhanced with the incorporation of 2D Contour filtering step before proceeding for 3D STL Model reconstruction, compared to the algorithm presented in \cite{wang2010stl}. 
Processing starts from bottom layer and successive layer 2D Contour data points. Delaunay Triangulation is performed on the 2D contour points of two layers. The Mathworks definition \cite{MATLAB2026} of delaunay triangulation is given below:

“Delaunay Triangulation: delaunay creates a Delaunay triangulation of a set of points in 2-D or -D space. A 2-D Delaunay triangulation ensures that the circumcircle associated with each triangle contains no other point in its interior.” 
Next a wall layer consisting of polygons is constructed between these two layers. The construction of polygon wall layer depends on the number of contour data points of each layer. There arise, two different cases as presented in \cite{wang2010stl} and are given below.

\begin{figure}[h]
\centering
\includegraphics[width=.5\textwidth]{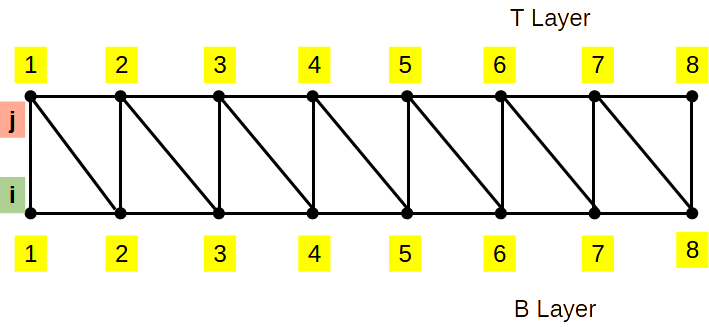}
\caption{Delaunay triangulation of adjacent contours with same number of data points.}
\label{fig:triangulation}
\end{figure}

\textbf{Case 1}: The number of 2D contour data points are same in each adjacent layer.
Delaunay Triangulation for two adjacent layers T and B with j and i data points is shown in Figure~\ref{fig:triangulation}. 
whenever, i is incremented, a triangle is added between the layers, and j is incremented. After, i , reaches the last data point, the triangulation is closed with starting point with the addition of new triangles. In Figure~\ref{fig:triangulation}, in T layer, no. of data points j=8  and S layer, no. of data points i = 8. After Delaunay Triangulation process reaches i=8, Triangulation is closed by connecting i=8 \& j =8 with i=1 \& j=1.  The total number of Delaunay Triangles are i+j = 8+8 =16.  

\textbf{Case 2}: The number of 2D contour data points are different in each adjacent layer.
Difference in points between two layers complicates the process of delaunay triangulation. In this work layer with most no. of points j is assigned to layer T and the one with less no. of points i to layer B.  The data points j and i are inputted to Equation~\ref{eq:eq2} and Equation~\ref{eq:eq3} for further processing.

\begin{equation}
d = j-i
\label{eq:eq2}
\end{equation}

and

\begin{equation}
\frac{d}{j} = q...r
\label{eq:eq3}
\end{equation}

where d is the point difference between layers j and i. q is the quotient representing the point requiring and extra triangulation. r is the remainder, representing the no. of data points, which are not considered for adding additional triangles. The values q and r can be further separated in to three categories. (i) Fully divisible, (ii) not fully divisible, with a remainder of r = 1, (iii) not fully divisible with a remainder  $r > 1$. When r = 0, it can be fully divided. When $r>0$, it can not be fully divided. We can then distinguish whether the r value is equal to 1 or greater than 1. After recognition using the above equations, delaunay triangulation construction can be separated in to three methods and are described in \cite{wang2010stl}. The above cases need to handled before proceeding for constructing the wall layer. The entire STL Mesh generation process is shown in Flowchart in Figure~\ref{fig:flowchart}.

\section{Simulation Results}
        \label{sec:simulation}
        To perform 2D Contour Data Points  to 3D STL Mesh reconstruction [33], we conducted simulation  using hp probook 64-bit, 8 GB RAM  installed with MATLAB R2014b. The 2D Contour Data points to 3D STL Mesh reconstruction  algorithm is implemented in MATLAB. The simulation is performed to reconstruct 3D Models of basic shapes and 2D Contour data points extracted from CT images and is presented in Case Study – I and Case Study – II. 

\subsection{Basic Shapes: Case Study I}
\begin{figure}[t]
\centering
\includegraphics[width=.40\textwidth]{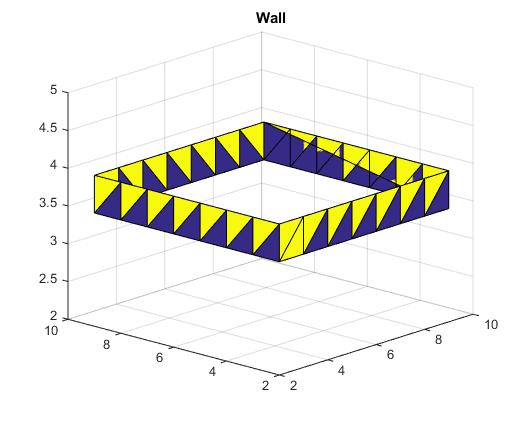}
\caption{3D STL Model of hollow Square.}
\label{fig:square}
\end{figure}

In this simulation clean 2D Point data set of basic shapes, hollow square and hollow circular ring are considered to reconstruct 3D STL Mesh  The 2D data points of hollow rectangle, layer 1  and layer 2 separated by a height of 1 unit are considered for wall layer construction. The wall layer constructed between layer 1 and layer 2  is shown in Figure~\ref{fig:square}  for hollow square.

\begin{figure}[t]
\centering
\includegraphics[width=.40\textwidth]{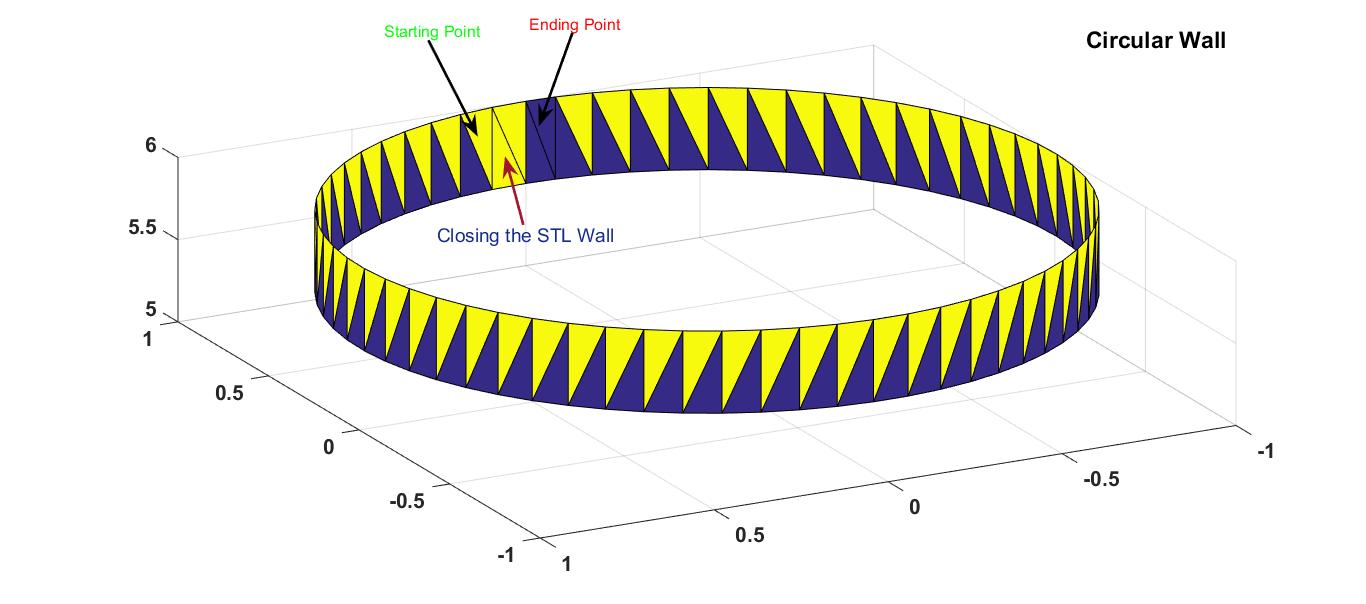}
\caption{3D STL Model of hollow Circle.}
\label{fig:circle}
\end{figure}

The constructed wall layer for 2D circular data points of layer-1 and layer-2 is shown in Figure~\ref{fig:circle}. In the figure the starting point and ending point are closed with additional triangle and indicated as “Closing the STL Wall”.

After the construction of wall layer between the 2D data points of two layers of basic shapes, 3D Mesh generation for three layers from 2D Contours is performed based on the \textbf{cases} presented in Section \ref{sec:method}. 
Total no. of layers considered three. l1, l2, l3
The distance between the successive layers is ‘z’
To construct a 3D mesh model from the delaunay triangulated layers of l1,  l2 and l3, two wall layers are required, The first wall layer is between l1 and l2. The second wall layer is between l2 and l3. The wall layers are constructed considering the three cases.

\begin{figure}[t]
\centering
\includegraphics[width=.40\textwidth]{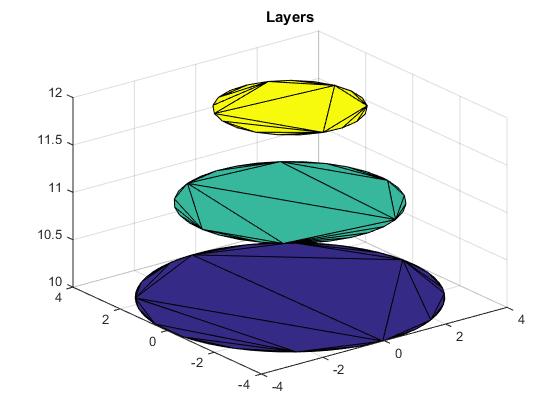}
\caption{Delaunay Triangulation of three 2D Layers l1, l2, l3.}
\label{fig:2Dlayers}
\end{figure}

\begin{figure}[t]
\centering
\includegraphics[width=.40\textwidth]{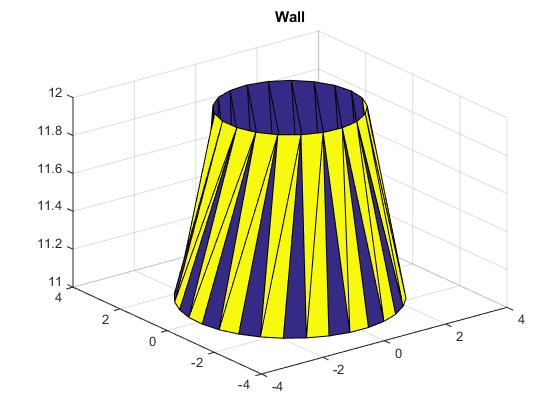}
\caption{Wall layer for layer1 and layer 2}
\label{fig:walllayers12}
\end{figure}

\begin{figure}[t]
\centering
\includegraphics[width=.40\textwidth]{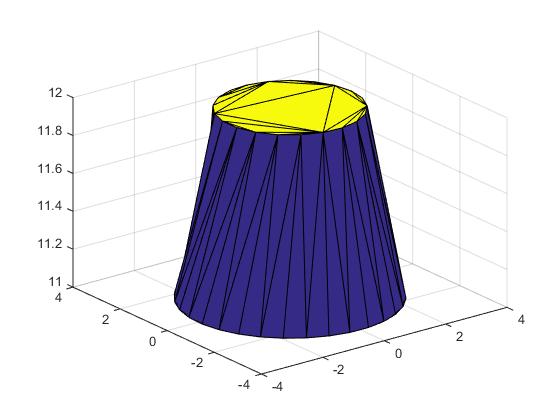}
\caption{3D Model of layer 1 and layer 2.}
\label{fig:3Dmodel12}
\end{figure}

\begin{figure}[t]
\centering
\includegraphics[width=.40\textwidth]{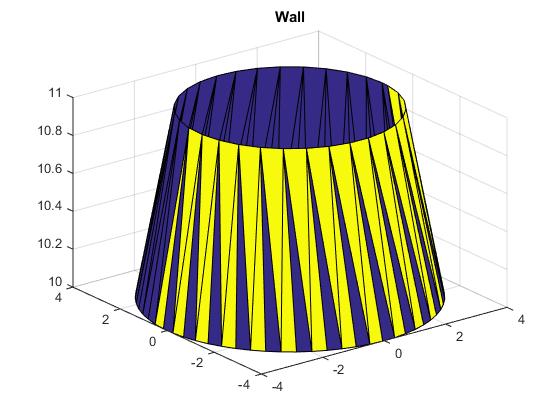}
\caption{Wall layer for layer2 and layer3.}
\label{fig:walllayer23}
\end{figure}

\begin{figure}[t]
\centering
\includegraphics[width=.40\textwidth]{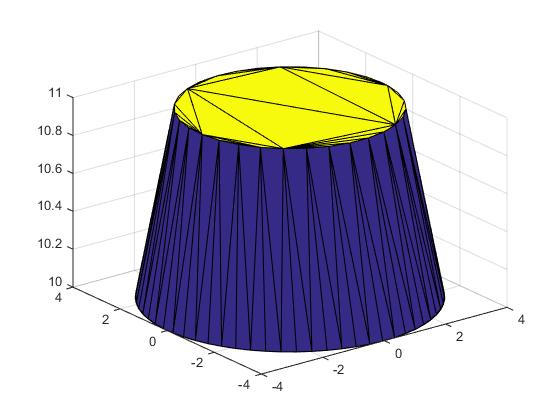}
\caption{3D Model of layer2 and layer3.}
\label{fig:model23}
\end{figure}

\begin{figure}[t]
\centering
\includegraphics[width=.40\textwidth]{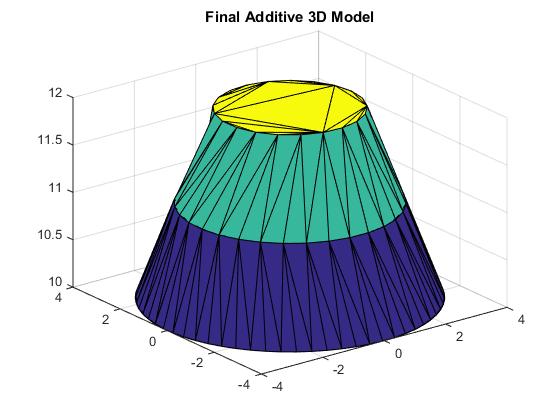}
\caption{3D STL Model for three layers.}
\label{fig:3Dmodel}
\end{figure}

First the 2D Contour points of l1,, l2, l3  are read in to MATLAB. Delaunay triangulation is performed on the 2D Contour points of l1, l2, l3 and is shown in Figure~\ref{fig:2Dlayers}.  The delaunay triangulation of wall layer constructed between layers l1 and l2 is shown in Figure~\ref{fig:walllayers12}. Delaunay triangulation of layer 1, Wall layer and layer 2 are merged and is shown in Figure~\ref{fig:3Dmodel12}. Similarly, delaunay triangulation of wall layer is constructed between layers l2 and l3 is shown in Figure~\ref{fig:walllayer23}. and merged layer 2, wall layer and layer 3 is shown in Figure~\ref{fig:model23}. The final 3D STL Delaunay triangulation model from 2D Points of three layers is obtained by combining the delaunay triangulated layers, wall layers and is shown in Figure~\ref{fig:3Dmodel}.
\textit{The matlab code to generate the 3D STL Model of the considered shape is deposited in Matlab File exchange with file ID – 83543.}      

\begin{figure}[t]
\centering
\includegraphics[width=.20\textwidth]{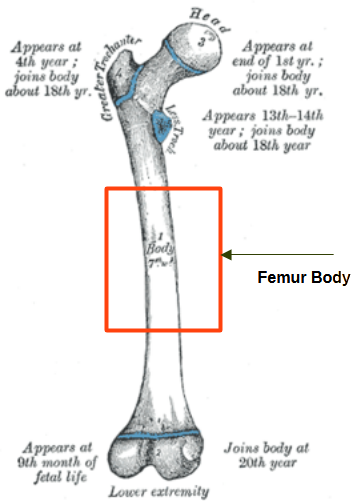}
\caption{Left femur seen from behind.}
\label{fig:femur1}
\end{figure}

\begin{figure*}[t]
\centering
\includegraphics[width=\textwidth]{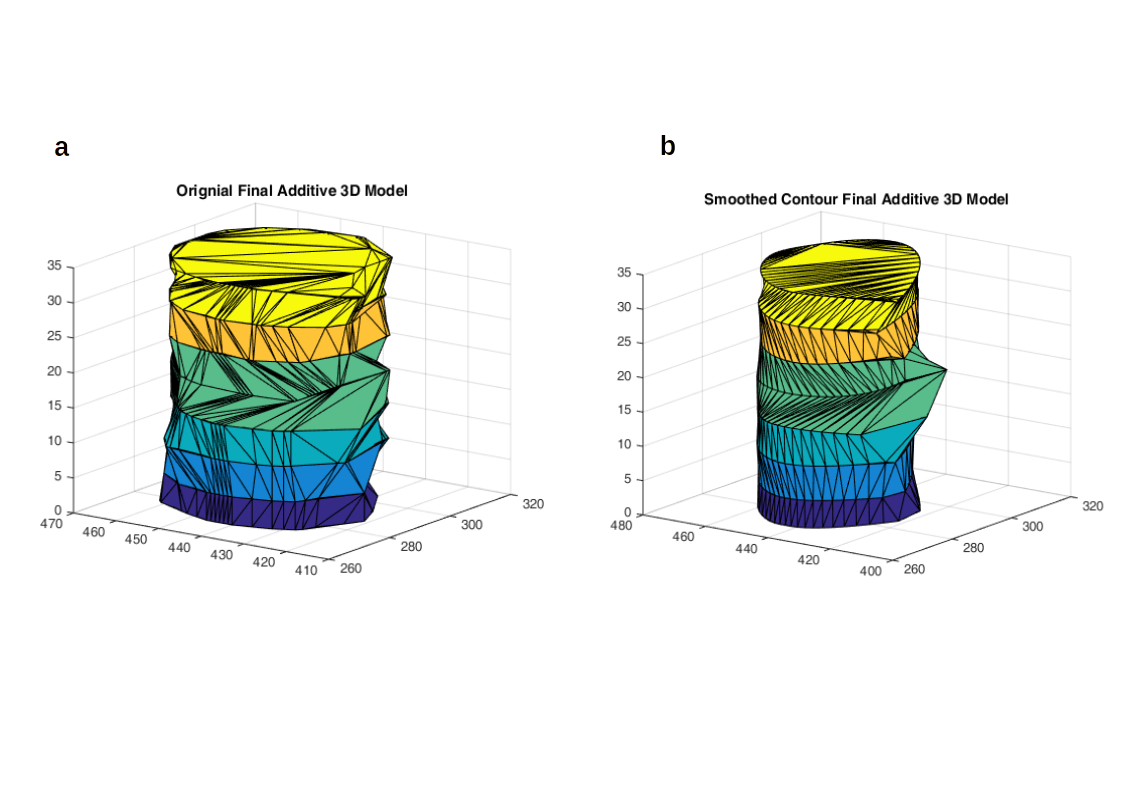}
\caption{a) without filtering, b) with filtering}
\label{fig:femurmodel}
\end{figure*}

\subsection{CT Images: Case Study II}
The CT data set consisting for femur bone for case study is considered from Osirix database \cite{databaseKey}. The human femur bone is shown in fig~\ref{fig:femur1}. First the CT images in DICOM format are read into Matlab. The details of the CT image are given in Table~\ref{tab:CTImagesParameters}.

\begin{table*}[t]
	\centering
		\begin{tabular}{|c|c|}
		\hline 
        \textbf{Parameter} & \textbf{Value} \\ 
        \hline 
        Height x Width & 512 x 512 \\
				\hline
        Bit Depth & 12 bits \\
				\hline
        Slice thickness & 1.5 mm\\ 
        \hline 
		\end{tabular}
	\caption{CT Images Parameters}
	\label{tab:CTImagesParameters}
\end{table*}

CT Images are visualized in Matlab and region of interest slices i.e, femur body is identified for 3D STL Model reconstruction. In this study CT Slices 80 to 87 are considered, as the ROI is present in these images. Second, the CT images are pre-processed in sequence using traditional image processing. The processed images are segmented to extract the 2D Contours of the ROI. Next, the 2D Contours are filtered using Moving average filter with span=0.1. The filtered 2D Data points are considered for the 3D STL Model reconstruction work flow as shown in Figure~\ref{fig:flowchart}. The wall thickness considered is 5mm. The reconstructed 3D STL  model from 2D contour data points extracted from slice 80 to 87 without filtering and with filtering is shown in Figure~\ref{fig:femurmodel} at viewing angle coordinates, Az: -55 and El: 18.  From the Figure~\ref{fig:femurmodel} it is subjectively clear that the filtered 2D Contour points represented the geometry of the segmented femur body.

\section{CONCLUSION}
        \label{sec:conclusion}
        CT images provide details in 2D image format and the clinician interprets them to identify any deformity or abnormality condition present in the internal structure of human body. To aid clinician in better understanding of CT images, the internal structure scanned need to be presented in 3D Model. In this work, an enhanced frame work is proposed for reconstructing 3D STL model to basic shapes and CT images. The CT images are first processed in Traditional Image processing work flow to make them suitable for segmentation task. 2D Contours of ROI are extracted from segmented CT images. The 2D contours are subjected to an additional filtering processing step with suitable span value to mitigate the segmentation errors and to aid in smooth reconstruction of 3D STL Model. Reconstruction results are presented without filtering and with filtering from extracted 2D Data points from segmented CT slice. The filtering step in the 3D STL mesh reconstruction resulted in smooth geometry of the model compared to the one without filtering. 

\section{ACKNOWLEDGEMENT}
        \label{sec:acknowledgement}
        This project was supported by the Kakatiya University.

\bibliography{main}

@article{young2008efficient,
  title={An efficient approach to converting three-dimensional image data into highly accurate computational models},
  author={Young, PG and Beresford-West, TBH and Coward, SRL and Notarberardino, B and Walker, B and Abdul-Aziz, A},
  journal={Philosophical Transactions of the Royal Society A: Mathematical, Physical and Engineering Sciences},
  volume={366},
  number={1878},
  pages={3155--3173},
  year={2008},
  publisher={The Royal Society London}
}

@book{preim2007visualization,
  title={Visualization in medicine: theory, algorithms, and applications},
  author={Preim, Bernhard and Bartz, Dirk},
  year={2007},
  publisher={Elsevier}
}

@article{karatas2014three,
  title={Three-dimensional imaging techniques: A literature review},
  author={Karatas, Orhan Hakki and Toy, Ebubekir},
  journal={European journal of dentistry},
  volume={8},
  number={01},
  pages={132--140},
  year={2014},
  publisher={Thieme Medical and Scientific Publishers Private Ltd.}
}

@article{ahn2006rapid,
  title={Rapid prototyping and reverse engineering application for orthopedic surgery planning},
  author={Ahn, Dong-Gyu and Lee, Jun-Young and Yang, Dong-Yol},
  journal={Journal of mechanical science and technology},
  volume={20},
  number={1},
  pages={19--28},
  year={2006},
  publisher={Springer}
}

@article{willis2007rapid,
  title={Rapid prototyping 3D objects from scanned measurement data},
  author={Willis, Andrew and Speicher, Jasper and Cooper, David B},
  journal={Image and Vision Computing},
  volume={25},
  number={7},
  pages={1174--1184},
  year={2007},
  publisher={Elsevier}
}

@article{marovic1998visualization,
  title={Visualization of 3D fields and medical data and using VRML},
  author={Marovic, Branko and Jovanovic, Zoran},
  journal={Future Generation Computer Systems},
  volume={14},
  number={1-2},
  pages={33--49},
  year={1998},
  publisher={Elsevier}
}

@inproceedings{pflesser1995towards,
  title={Towards realistic visualization for surgery rehearsal},
  author={Pflesser, Bernhard and Tiede, Ulf and H{\"o}hne, Karl Heinz},
  booktitle={International Conference on Computer Vision, Virtual Reality, and Robotics in Medicine},
  pages={487--491},
  year={1995},
  organization={Springer}
}

@article{rosa2004rapid,
  title={Rapid prototyping in maxillofacial surgery and traumatology},
  author={Rosa, Everton Luis Santos da and Oleskovicz, C{\'e}sar Fernando and Aragao, Bruno Nogueira},
  journal={Brazilian Dental Journal},
  volume={15},
  pages={243--247},
  year={2004},
  publisher={SciELO Brasil}
}

@article{wang2010stl,
  title={STL rapid prototyping bio-CAD model for CT medical image segmentation},
  author={Wang, Chung-Shing and Wang, Wei-Hua A and Lin, Man-Ching},
  journal={Computers in Industry},
  volume={61},
  number={3},
  pages={187--197},
  year={2010},
  publisher={Elsevier}
}

@article{manmadhachary2016improve,
  title={Improve the accuracy, surface smoothing and material adaption in STL file for RP medical models},
  author={Manmadhachary, A and Kumar, Ravi and Krishnanand, L},
  journal={Journal of Manufacturing Processes},
  volume={21},
  pages={46--55},
  year={2016},
  publisher={Elsevier}
}

@article{negi2014basics,
  title={Basics and applications of rapid prototyping medical models},
  author={Negi, Sushant and Dhiman, Suresh and Kumar Sharma, Rajesh},
  journal={Rapid Prototyping Journal},
  volume={20},
  number={3},
  pages={256--267},
  year={2014},
  publisher={Emerald Group Publishing Limited}
}

@article{ma2001rapid,
  title={Rapid prototyping applications in medicine. Part 1: NURBS-based volume modelling},
  author={Ma, D and Lin, F and Chua, CK},
  journal={The International Journal of Advanced Manufacturing Technology},
  volume={18},
  number={2},
  pages={103--117},
  year={2001},
  publisher={Springer}
}

@article{rajon2003marching,
  title={Marching cube algorithm: review and trilinear interpolation adaptation for image-based dosimetric models},
  author={Rajon, Didier A and Bolch, Wesley E},
  journal={Computerized Medical Imaging and Graphics},
  volume={27},
  number={5},
  pages={411--435},
  year={2003},
  publisher={Elsevier}
}

@book{choi1991surface,
  title={Surface modeling for CAD-CAM},
  author={Choi, Byoung K},
  year={1991},
  publisher={Elsevier Science Inc.}
}

@article{lee2002feature,
  title={Feature-guided shape-based image interpolation},
  author={Lee, Tong-Yee and Lin, Chao-Hung},
  journal={IEEE transactions on medical imaging},
  volume={21},
  number={12},
  pages={1479--1489},
  year={2002},
  publisher={IEEE}
}

@article{meyers1992surfaces,
  title={Surfaces from contours},
  author={Meyers, David and Skinner, Shelley and Sloan, Kenneth},
  journal={ACM Transactions On Graphics (TOG)},
  volume={11},
  number={3},
  pages={228--258},
  year={1992},
  publisher={ACM New York, NY, USA}
}

@article{shang2018closed,
  title={Closed T-spline surface reconstruction from medical image data},
  author={Shang, Ce and Fu, Jianzhong and Lin, Zhiwei and Feng, Jiawei and Li, Bin},
  journal={International Journal of Precision Engineering and Manufacturing},
  volume={19},
  number={11},
  pages={1659--1671},
  year={2018},
  publisher={Springer}
}

@article{maksimovic2000computed,
  title={Computed tomography image analyzer: 3D reconstruction and segmentation applying active contour models—‘snakes’},
  author={Maksimovic, Ruzica and Stankovic, Srdjan and Milovanovic, Dragorad},
  journal={International journal of medical informatics},
  volume={58},
  pages={29--37},
  year={2000},
  publisher={Elsevier}
}

@article{choi1988triangulation,
  title={Triangulation of scattered date in 3D space},
  author={Choi, BK and Shin, HY and Yoon, Y Ietal and Lee, JW},
  journal={Computer-aided design},
  volume={20},
  number={5},
  pages={239--248},
  year={1988},
  publisher={Elsevier}
}

@article{schumaker1993triangulations,
  title={Triangulations in CAGD},
  author={Schumaker, Larry L},
  journal={IEEE Computer Graphics and Applications},
  volume={13},
  number={1},
  pages={47--52},
  year={1993},
  publisher={IEEE}
}

@article{sisias2002algorithms,
  title={Algorithms for accurate rapid prototyping replication of cancellous bone voxel maps},
  author={Sisias, G and Phillips, R and Dobson, CA and Fagan, MJ and Langton, CM},
  journal={Rapid Prototyping Journal},
  volume={8},
  number={1},
  pages={6--24},
  year={2002},
  publisher={MCB UP Ltd}
}

@article{lin2007voxelization,
  title={Voxelization and fabrication of freeform models},
  author={Lin, F and Seah, HS and Wu, Z and Ma, D},
  journal={Virtual and Physical Prototyping},
  volume={2},
  number={2},
  pages={65--73},
  year={2007},
  publisher={Taylor \& Francis}
}

@article{leong1996study,
  title={A study of stereolithography file errors and repair. Part 1. Generic solution},
  author={Leong, KF and Chua, CK and Ng, YM},
  journal={The International Journal of Advanced Manufacturing Technology},
  volume={12},
  number={6},
  pages={407--414},
  year={1996},
  publisher={Springer}
}

@article{stroud2000stl,
  title={STL and extensions},
  author={Stroud, I and Xirouchakis, Paul C},
  journal={Advances in Engineering Software},
  volume={31},
  number={2},
  pages={83--95},
  year={2000},
  publisher={Elsevier}
}

@article{wu2006enhanced,
  title={Enhanced stl},
  author={Wu, Tong and Cheung, Edmund HM},
  journal={The International Journal of Advanced Manufacturing Technology},
  volume={29},
  number={11},
  pages={1143--1150},
  year={2006},
  publisher={Springer}
}

@article{melchels2010review,
  title={A review on stereolithography and its applications in biomedical engineering},
  author={Melchels, Ferry PW and Feijen, Jan and Grijpma, Dirk W},
  journal={Biomaterials},
  volume={31},
  number={24},
  pages={6121--6130},
  year={2010},
  publisher={Elsevier}
}

@misc{exampleWebsite,
  author       = {Radiopaedia},
  title        = {Hounsfield},
  year         = {2024},
  howpublished = {\url{https://radiopaedia.org/articles/hounsfield-unit}},
  note         = {Accessed: 2024-07-22}
}

@misc{databaseKey,
  author       = {Pixmeo SARL},
  title        = {Pelvix},
  year         = {2021},
  howpublished = {\url{https://www.osirix-viewer.com/}},
  note         = {Accessed: 2021-01-9}
}

@inproceedings{wang2013feature,
  title={Feature reconstruction for 3D medical images processing},
  author={Wang, Chung-Shing and Lin, Man-Ching and Wang, Chung-Chuan and Chen, Ching-Fu and Hsieh, Jei-Chen},
  booktitle={2013 6th International Conference on Biomedical Engineering and Informatics},
  pages={69--74},
  year={2013},
  organization={IEEE}
}

@inproceedings{chandar2016segmentation,
  title={Segmentation and 3D visualization of pelvic bone from CT scan images},
  author={Chandar, K Punnam and Satyasavithri, T},
  booktitle={2016 IEEE 6th International Conference on Advanced Computing (IACC)},
  pages={430--433},
  year={2016},
  organization={IEEE}
}

@manual{MATLAB2026,
  title        = {MATLAB},
  author       = {The MathWorks, Inc.},
  organization = {MathWorks},
  address      = {Natick, Massachusetts},
  year         = {2026},
  note         = {Version R2026a},
  url          = {https://www.mathworks.com/products/matlab.html}
}

@article{kanumilli2024advancements,
  title={Advancements and applications of three-dimensional printing technology in surgery},
  author={Kanumilli, Sri Lakshmi Devi and Kosuru, Bhanu P and Shaukat, Faiza and Repalle, Uday Kumar},
  journal={Journal of Medical Physics},
  volume={49},
  number={3},
  pages={319--325},
  year={2024},
  publisher={Medknow}
}
\bibliographystyle{IEEEtran}

\end{document}